\documentclass[aps,prc,preprint,amsmath,amssymb,showpacs,preprintnumbers,superscriptaddress]{revtex4-1}
\usepackage{CJK}
\usepackage{graphicx}
\usepackage{dcolumn}
\usepackage{bm}
\usepackage{color}
\usepackage{hyperref}

\allowdisplaybreaks

\begin{document}
\begin{CJK*}{UTF8}{}
\title{Generalized time-dependent generator coordinate method for small and large amplitude collective motion}
\CJKfamily{gbsn}
\author{B. Li}
\affiliation{State Key Laboratory of Nuclear Physics and Technology, School of Physics, Peking University, Beijing 100871, China}
\author{D. Vretenar}
\email{vretenar@phy.hr}
\affiliation{Physics Department, Faculty of Science, University of Zagreb, 10000 Zagreb, Croatia}
\affiliation{State Key Laboratory of Nuclear Physics and Technology, School of Physics, Peking University, Beijing 100871, China}
\author{T. Nik\v si\' c}
\affiliation{Physics Department, Faculty of Science, University of Zagreb, 10000 Zagreb, Croatia}
\affiliation{State Key Laboratory of Nuclear Physics and Technology, School of Physics, Peking University, Beijing 100871, China}
\author{P. W. Zhao}
\email{pwzhao@pku.edu.cn}
\affiliation{State Key Laboratory of Nuclear Physics and Technology, School of Physics, Peking University, Beijing 100871, China}
\author{J. Meng}
\email{mengj@pku.edu.cn}
\affiliation{State Key Laboratory of Nuclear Physics and Technology, School of Physics, Peking University, Beijing 100871, China}

\begin{abstract}
An implementation of the generalized time-dependent generator coordinated method (TD-GCM) is developed, that can be applied to the dynamics of small- and large-amplitude collective motion of atomic nuclei. Both the generator states and weight functions of the GCM correlated wave function depend on time. The initial generator states are obtained as solutions of deformation-constrained self-consistent mean-field equations, and are evolved in time by the standard mean-field equations of nuclear density functional theory (TD-DFT). The TD-DFT trajectories are used as a generally non-orthogonal and overcomplete basis in which the TD-GCM wave function is expanded. The weights, expressed in terms of a collective wave function, obey a TD-GCM (integral) equation. In this explorative paper, the generalized TD-GCM is applied to the excitation energies and spreading width of giant resonances, and to the dynamics of induced fission. The necessity of including pairing correlations in the basis of TD-DFT trajectories is demonstrated in the latter example.
\end{abstract}

\date{\today}

\maketitle

\end{CJK*}
\section{Introduction}
Two basic microscopic frameworks have been used in the last decade for a quantitative analysis of collective time-dependent processes in atomic nuclei. The first one includes a number of models based on nuclear time-dependent density functional theory (TD-DFT) \cite{simenel12,simenel18,nakatsukasa16,stevenson19,bulgac16,magierski17,scamps18,bulgac19,bulgac20,Ren2022}. Given a nuclear energy density functional (EDF) and pairing interaction, TD-DFT can be used to model a variety of complex phenomena, from small-amplitude collective oscillations of the nuclear density, to large-amplitude processes such as fission and heavy-ion reactions. However, since the TD-DFT-based model describes the classical evolution of independent nucleons in mean-field potentials, that is, the propagation of the one-body density, it cannot describe the spreading widths of one-body observables \cite{Reinhard1983NPA}. In the case of nuclear fission, in particular, TD-DFT automatically includes the one-body dissipation mechanism, but can only simulate a single fission event by propagating the nucleons independently.
Therefore, even though this approach has been very successful when calculating the total kinetic energy in the fission process, it cannot be used for a quantitative estimate of the widths of charge or mass fragment distributions.

In the time-dependent generator coordinate method (TD-GCM) \cite{krappe12,schunck16,younes19,Regnier2016_PRC93-054611,Verriere2020_FP8-233}, the nuclear wave function is expressed in terms of a superposition of generator states that are functions of collective coordinates. GCM presents a fully quantum mechanical approach but only takes into account collective degrees of freedom, such as shape variables and pairing degrees of freedom. For time-dependent phenomena, it has mostly been applied in the adiabatic Gaussian overlap approximation, in which a Schr\"odinger-like equation governs the time evolution of the nuclear wave function in the space of collective coordinates. The dissipation of the energy of collective motion into intrinsic degrees of freedom plays an important role in the description of collective dynamics and, therefore, for a quantitative modeling of time-dependent processes, it is necessary to expand the TD-GCM beyond the adiabatic approximation. Several microscopic extensions of TD-GCM that include diabatic effects have been considered \cite{dietrich10,bernard11,younes19}, but they are rather complex and have yet to be implemented in a model that is realistic from a computational point of view.
In two recent studies \cite{Zhao_PRC105,Zhao_PRC106}, the TD-GCM has been extended to allow for dissipation effects in the description of induced fission dynamics. The framework is based on the quantum theory of dissipation for nuclear collective motion \cite{Kerman1974_PS10-118},
and introduces a generalization of the GCM generating functions that includes excited states, and the resulting equation of motion in the collective coordinates and excitation energy. In the case of induced fission, and with a single phenomenological strength parameter of the dissipation term, the model provides a quantitative description of both the fission yields and total kinetic energy distributions.

Another possibility is to directly combine the TD-GCM and TD-DFT, in an approach that preserves the quantum mechanical description of collective dynamics intrinsic to the GCM, and at the same time extends the GCM beyond the adiabatic approximation. Here we adopt a method that was originally introduced in Ref.~\cite{Reinhard1983NPA}, but fully developed only more recently and applied to particle number restoration in a description of transfer of particles between two finite superfluid systems \cite{Regnier2019PRC}, and to collective multi-phonon states in nuclei \cite{Marevie2023}. In this method, both the generator states and weight functions of the GCM correlated wave function depend on time. The initial generator states are obtained as solutions of deformation-constrained self-consistent mean-field equations, and are evolved in time by the standard mean-field equations of nuclear density functional theory (TD-DFT). The TD-DFT trajectories are used as a generally non-orthogonal and overcomplete basis in which the TD-GCM wave function is expanded. The weights, expressed in terms of a collective wave function, obey a time-dependent GCM (integral) equation.

In Sec.~\ref{sec_theo} we develop the theoretical framework, and specialize to the particular model that will be used in the  present and future calculations. Section~\ref{sec:208Pb} presents an illustrative study of small-amplitude collective vibrations in $^{208}$Pb. The response to the monopole, quadrupole, octupole, and hexadecapole operators is analyzed, and it is shown that the inclusion of mode coupling in the generalized TD-GCM is necessary for the description of the spreading widths of resonances. As a simple example of large-amplitude motion, in Sec.~\ref{sec_fission} a schematic TD-GCM description of induced fission dynamics of $^{240}$Pu is discussed. Section~\ref{sec_summ} summarizes the results and presents a brief outlook for future studies. In the Appendix we include (a) an example of monopole oscillations of  $^{16}$O for which, in the case of a non-orthogonal and overcomplete basis of TD-DFT trajectories, it is necessary to project, at each time step, the eigenvectors of the overlap kernel with negligible (close to zero) eigenvalues and (b) the details of the calculation of strength functions.

\section{Theoretical framework: Generalized time-dependent GCM}\label{sec_theo}

In the framework of the generalized TD-GCM, the Griffin-Hill-Wheeler  (GHW) ansatz for the correlated nuclear wave function reads~\cite{Reinhard1983NPA,Regnier2019PRC,Verriere2020_FP8-233}
\begin{equation}
    |\Psi(t)\rangle=\int_{\bm q} d{\bm q}~f_{\bm q}(t) |\Phi_{\bm q}(t)\rangle,
    \label{GHW_wf}
\end{equation}
where the vector ${\bm q}$ denotes the continuous real {\em generator coordinates} that parametrize, for instance, the shape of the nucleus. The nuclear wave function is a linear superposition of, generally non-orthogonal, many-body {\em generator states} $|\Phi_{\bm q}(t)\rangle$, and $f_{\bm q}(t)$ are the corresponding complex-valued {\em weight functions}. In the static case, in which the GCM is used to calculate excitation spectra or restore broken symmetries, neither the weight functions nor the generator states depend on time. In most time-dependent applications, only the weights $f_{\bm q}(t)$ are functions of time, while the generator states $|\Phi_{\bm q}\rangle$ are usually solutions of constrained Hartree-Fock-Bogoliubov (HFB) calculations, with constraints on the mass multipole moments. This choice of generator states corresponds to the adiabatic approximation, because their energy is minimized under certain constraints, and remains such during the time evolution of the system. The equation of motion of the TD-GCM yields only the probability that the system will occupy these predefined states \cite{Verriere2020_FP8-233}, and does not include any dissipation mechanism.

In the generalized TD-GCM the generator states $|\Phi_{\bm q}(t)\rangle$ are determined dynamically, starting from some initial conditions. If specific constrained-HFB solutions are selected as initial conditions, $|\Phi_{\bm q}(t)\rangle$ are Slater determinants that obey the TD-DFT (TD-HFB) equations, that is, they describe the classical evolution of independent nucleons in self-consistent mean-field potentials and correspond to trajectories in the space of collective coordinates. The TD-DFT automatically includes the one-body dissipation mechanism and, therefore, by combining the Slater determinants $|\Phi_{\bm q}(t)\rangle$ with the variationally determined equation of motion for the weight functions $f_{\bm q}(t)$, one obtains a fully quantum mechanical description of collective dynamics that goes beyond the adiabatic approximation and includes quantum fluctuations.

The model we employ in this explorative paper parallels the method developed in Ref.~\cite{Regnier2019PRC}. For simplicity, pairing correlations are not taken into account, and the discretized generator coordinates are the mass multipole moments (monopole, quadrupole, octupole, and hexadecapole) of the nucleon density distribution. Thus, the nuclear wave function
\begin{equation}
    |\Psi(t)\rangle=\sum_{\bm q} f_{\bm q}(t) |\Phi_{\bm q}(t)\rangle.
    \label{Eq_collec_wfs}
\end{equation}
is the solution of the time-dependent equation
\begin{equation}
    i\hbar\partial_t|\Psi(t)\rangle=\hat{H}|\Psi(t)\rangle,
    \label{Eq_td_eq}
\end{equation}
where $\hat{H}$ is the Hamiltonian of the nuclear system. From a time-dependent variational principle~\cite{Regnier2019PRC}, one obtains the equation of motion for the weight functions
\begin{equation}
     i\hbar \mathcal{N}\dot{f}=(\mathcal{H}-\mathcal{H}^{MF})f,
    \label{TD-HW-f}
\end{equation}
which, in the discretized collective space, reads
\begin{equation}
\sum_{\bm q} i\hbar \mathcal{N}_{\bm{q'q}}(t)\partial_t f_{\bm q}(t)+\sum_q \mathcal{H}_{\bm {q'q}}^{MF}(t) f_{\bm q}(t) =\sum_q \mathcal{H}_{\bm {q'q}}(t)f_{\bm q}(t).
\end{equation}
The time-dependent kernels
\begin{subequations}
 \begin{align}
&\mathcal{N}_{\bm{q'q}}(t)=\langle\Phi_{\bm {q'}}(t)|\Phi_{\bm q}(t)\rangle,\label{Eq_N}\\
&\mathcal{H}_{\bm{q'q}}(t)=\langle\Phi_{\bm {q'}}(t)|\hat{H}|\Phi_{\bm q}(t)\rangle,\label{Eq_H}\\
&\mathcal{H}^{MF}_{\bm{q'q}}(t)=\langle\Phi_{\bm {q'}}(t)|i\hbar\partial_t|\Phi_{\bm q}(t)\rangle,
\label{Eq_H_mf}
 \end{align}
\end{subequations}
include the overlap, the Hamiltonian, and the time derivative of the generator states, respectively.
\subsection{Time-dependent Slater determinant $|\Phi_{\bf q}(t)\rangle$}

The time evolution of a Slater determinant characterized by a generator coordinate ${\bm q}$, for a nucleus with $A$ nucleons
\begin{equation}
   |\Phi_{\bm q}(t)\rangle = \prod_{k=1}^A c_{{\bm q},k}^+(t)|-\rangle,
\label{Eq_Slater}
\end{equation}
is modeled by the time-dependent covariant density functional theory~\cite{Ren2020PLB,Ren2020PRC}.
The initial Slater determinant $|\Phi_{\bm q}(t=0)\rangle$ is a solution of self-consistent mean-field equations, with constraints on the mass multipole moments of the nucleon density distribution. The corresponding single-particle states $\phi_k^{\bm q}(\bm{r},t)$ are solutions of the time-dependent Dirac equation
\begin{equation}\label{Eq_td_Dirac_eq_BCS}
  i\frac{\partial}{\partial t}\phi_k^{\bm q}(\bm{r},t)=\hat{h}^{\bm q}(\bm{r},t)\phi_k^{\bm q}(\bm{r},t),
\end{equation}
where the single-particle Hamiltonian $\hat{h}^{\bm q}(\bm{r},t)$ is given by
\begin{equation}
   \hat{h}^{\bm q}(\bm{r},t) = \bm{\alpha}\cdot(\hat{\bm{p}}-\bm{V}_{\bm q})+V^0_{\bm q}+\beta(m_N+S_{\bm q}).
   \label{Ham_D}
\end{equation}
Here, $m_N$~is the nucleon mass, and the scalar $S_{\bm q}(\bm{r},t)$ and four-vector $V^{\mu}_{\bm q}(\bm{r},t)$ potentials are determined by the time-dependent densities and currents in the isoscalar-scalar, isoscalar-vector, and isovector-vector channels. In this paper we employ the point-coupling relativistic energy density functional PC-PK1 \cite{Zhao2010PRC}, and the explicit expressions for the potentials read
\begin{subequations}
  \begin{align}
    S_{\bm q}(\bm{r})=\,&\alpha_S\rho_S^{\bm q}+\beta_S(\rho_S^{\bm q})^2+\gamma_S(\rho_S^{\bm q})^3+\delta_S\Delta\rho_S^{\bm q},\\
    V^\mu_{\bm q}(\bm{r})=\,&\alpha_Vj^{{\bm q},\mu}+\gamma_V(j^{{\bm q},\mu} j^{\bm q}_\mu)j^{{\bm q},\mu}+\delta_V\Delta j^{{\bm q},\mu}+\tau_3\alpha_{TV}j_{TV}^{{\bm q},\mu}+\tau_3\delta_{TV}\Delta j_{TV}^{{\bm q},\mu}+e\frac{1-\tau_3}{2}A^{{\bm q},\mu},
  \end{align}
\end{subequations}
where $\tau_3$ is the isospin Pauli matrix, and $A^{{\bm q},\mu}$ is the electromagnetic vector potential.
The densities and currents are defined in terms of occupied single-particle wave functions~$\phi_k^{\bm q}(\bm{r},t)$:
\begin{subequations}\label{Eq_density_current}
  \begin{align}
    &\rho_S^{\bm q}(\bm{r},t)=\sum_k^{A}\bar{\phi}^{\bm q}_k(\bm{r},t)\phi^{\bm q}_k(\bm{r},t),\\
    &j^{{\bm q},\mu}(\bm{r},t)=\sum_k^{A}\bar{\phi}^{\bm q}_k(\bm{r},t)\gamma^\mu\phi^{\bm q}_k(\bm{r},t),\\
    &j_{TV}^{{\bm q},\mu}(\bm{r},t)=\sum_k^{A}\bar{\phi}^{\bm q}_k(\bm{r},t)\gamma^\mu\tau_3\phi^{\bm q}_k(\bm{r},t).
  \end{align}
\end{subequations}

\subsection{Overlap kernel $\mathcal{N}_{\bf {q'q}}(t)$}
According to Eq.(\ref{Eq_Slater}), the expression for the overlap kernel Eq.(\ref{Eq_N}) can be written in the following form:
\begin{equation}\label{Eq_norm_kernel}
    \begin{split}
    \mathcal{N}_{\bm {q'q}}(t)&=\langle\Phi_{\bm {q'}}(t)|\Phi_{\bm q}(t)\rangle\\
    &=(-1)^{A(A-1)/2}~\langle-|c_{\bm {q'},1}(t)...c_{\bm {q'},A}(t)c^{\dagger}_{\bm q,1}(t)...c^{\dagger}_{\bm q,A}(t)|-\rangle. \\
    \end{split}
\end{equation}
The overlap between two Slater determinants can be calculated by the Pfaffian algorithms
proposed in Refs~\cite{Robledo2009PRC,Hu2014PLB}.

\subsection{Energy kernel $\mathcal{H}_{\bf {q'q}}(t)$}

For the point-coupling relativistic energy density functional PC-PK1 \cite{Zhao2010PRC},
one obtains the expression for the energy kernel $\mathcal{H}(t)$, under the assumption \cite{nakatsukasa16} that it only depends on the transition densities at time $t$:
\begin{equation}
\begin{aligned}
        \mathcal{H}_{\bm {q'q}}(t)=\langle\Phi_{\bm {q'}}(t)|\hat{H}|\Phi_{\bm q}(t)\rangle&=\langle\Phi_{\bm {q'}}(t)|\Phi_{\bm q}(t)\rangle\cdot\int d^3r~\{ \rho_{\rm kin}(\bm{r},t)\\
        &+\frac{\alpha_S}{2}\rho_S(\bm{r},t)^2+\frac{\beta_S}{3}\rho_S(\bm{r},t)^3\\
        &+\frac{\gamma_S}{4}\rho_S(\bm{r},t)^4+\frac{\delta_S}{2}\rho_S(\bm{r},t)\Delta\rho_S(\bm{r},t)\\
        &+\frac{\alpha_V}{2} j^\mu(\bm{r},t) j_\mu(\bm{r},t)+\frac{\gamma_V}{4}(j^\mu(\bm{r},t) j_\mu(\bm{r},t))^2\\
        &+\frac{\delta_V}{2} j^\mu(\bm{r},t)\Delta j_\mu(\bm{r},t)+\frac{\alpha_{TV}}{2} j^\mu_{TV}(\bm{r},t)\cdot [j_{TV}(\bm{r},t)]_\mu\\
        &+\frac{ \delta_{TV}}{2} j^\mu_{TV}(\bm{r},t)\cdot \Delta[j_{TV}(\bm{r},t)]_\mu+\frac{e^2}{2}j^\mu_p(\bm{r},t) A_\mu(\bm{r},t) \},
\end{aligned}
\end{equation}
where the densities and currents~$\rho_{\rm kin}$,~$\rho_S$,~$j^\mu$,~$j_{TV}^\mu$,~and~$j_p^\mu$ read
\begin{subequations}
 \begin{align}
        &\rho_{\rm kin}(\bm{r},t)=\sum_{l_1}^{A}\sum_{l_2}^{A}\bar{\phi}_{l_1}^{\bm {q'}}(\bm{r},t)(-i\bm{\gamma}\cdot\bm{\nabla}+m_N)\phi_{l_2}^{\bm q}(\bm{r},t)\rho^{\rm tran}_{l_1l_2}(t),\\
        &\rho_S(\bm{r},t)=\sum_{l_1}^{A}\sum_{l_2}^{A}\bar{\phi}_{l_1}^{\bm {q'}}(\bm{r},t)\phi_{l_2}^{\bm q}(\bm{r},t)\rho^{\rm tran}_{l_1l_2}(t),\\
        &j^\mu(\bm{r},t)=\sum_{l_1}^{A}\sum_{l_2}^{A}\bar{\phi}_{l_1}^{\bm {q'}}(\bm{r},t)\gamma^\mu\phi_{l_2}^{\bm q} (\bm{r},t)\rho^{\rm tran}_{l_1l_2}(t),\\
        &j_{TV}^\mu(\bm{r},t)=\sum_{l_1}^{A}\sum_{l_2}^{A}\bar{\phi}_{l_1}^{{\bm q'}}(\bm{r},t)\tau_3\gamma^\mu\phi_{l_2}^{\bm q}(\bm{r},t)\rho^{\rm tran}_{l_1l_2}(t),\\
        &j_p^\mu(\bm{r},t)=\frac{1-\tau_3}{2}\sum_{l_1}^{A}\sum_{l_2}^{A}\bar{\phi}_{l_1}^{\bm {q'}}(\bm{r},t)\gamma^\mu\phi_{l_2}^{\bm q}(\bm{r},t)\rho^{\rm tran}_{l_1l_2}(t).
\end{align}
\end{subequations}
The transition density matrix $\rho^{\rm tran}(t)$ is defined by the following relation
\begin{equation}
\rho^{\rm tran}_{l_1l_2}(t)=\frac{ \langle\Phi_{\bm {q'}}(t)|c_{{\bm {q'}},l_1}^{\dagger}(t)c_{{\bm q},l_2}(t)|\Phi_{\bm q}(t)\rangle}{\langle\Phi_{\bm {q'}}(t)|\Phi_{\bm q}(t)\rangle}.
\end{equation}
The numerator of the transition density matrix $\rho^{\rm
tran}_{l_1l_2}(t)$ is the overlap between two Slater determinants with $A-1$ particles. It can be calculated using the  Pfaffian algorithms~\cite{Hu2014PLB,Robledo2009PRC}.

\subsection{Mean-field kernel $\mathcal{H}^{MF}_{\bf {q'q}}(t)$}
From the expression for the time evolution of $|\Phi_{\bm q}(t)\rangle$~\cite{Ren2020PLB,Ren2020PRC},
\begin{equation}
   i\hbar\partial_t|\Phi_{\bm q}(t)\rangle=\sum_{l_2}^A\hat{h}^{\bm q}(\bm{r},t)c_{{\bm q},l_2}^{\dagger}(t)c_{{\bm q},l_2}(t)|\Phi_{\bm q}(t)\rangle,
\end{equation}
Eq.(\ref{Eq_H_mf}) can be written in the form
\begin{equation}
\mathcal{H}^{MF}_{\bm {q'q}}(t)=\langle\Phi_{\bm q'}(t)|i\hbar\partial_t|\Phi_{\bm q}(t)\rangle=\langle\Phi_{\bm q'}(t)|\sum_{l_2}^A\hat{h}^{\bm q}(\bm{r},t)c_{{\bm q},{l_2}}^{\dagger}(t)c_{{\bm q},{l_2}}(t)|\Phi_{\bm q}(t)\rangle.
\end{equation}
By expanding $\hat{h}^{\bm q}(\bm{r},t)c^{\dagger}_{{\bm q},l_2}(t)$ in a complete basis $ c^{\dagger}_{{\bm q'},l_1}(t)$,
\begin{equation}
     \hat{h}^{\bm q}(\bm{r},t)c^{\dagger}_{{\bm q},l_2}(t)=\sum_{l_1}\langle \phi^{{\bm q'}}_{l_1}(\bm{r},t)|\hat{h}^{\bm q}(\bm{r},t)|\phi^{\bm q}_{l_2}(\bm{r},t)\rangle c^{\dagger}_{{\bm q'},l_1}(t),
\end{equation}
one obtains for $\mathcal{H}_{\bm {q'q}}^{MF}(t)$ the expression
\begin{equation}
\mathcal{H}_{\bm{q'q}}^{MF}(t)=\langle\Phi_{\bm q'}(t)|\Phi_{\bm q}(t)\rangle\cdot\sum_{l_1}^{A}\sum_{l_2}^A\langle \phi^{\bm q'}_{l_1}(\bm{r},t)|\hat{h}^{\bm q}(\bm{r},t)|\phi^{\bm q}_{l_2}(\bm{r},t)\rangle\rho^{\rm tran}_{l_1l_2}(t).
\end{equation}

\subsection{Projection of spurious solutions (symmetric orthogonalization)}

 Because the basis of generator states $\Phi_{\bm q}(t)$ is generally non-orthogonal and overcomplete,
 it is necessary to remove the eigenvectors of the overlap kernel with negligible (close to zero) eigenvalues that preclude the inversion of the matrix $\mathcal{N}^{1/2}$ \cite{Regnier2019PRC}. This is performed by diagonalizing the overlap kernel $\mathcal{N}$:
\begin{equation}
     \mathcal{N}= \mathcal{U} \mathcal{D} \mathcal{U}^{\dagger}\rightarrow  \mathcal{N}_{{\bm {qq'}}}=\sum_{l} n_l\mathcal{U}_{{\bm q}l}\mathcal{U}^{\dagger}_{l{\bm q'}}
\end{equation}
where~$\mathcal{D}$~is the diagonal matrix of eigenvalues $n_l$ of the overlap kernel, and the columns of $\mathcal{U}$ form an orthonormal eigenbasis.

Next, a projection operator $\mathcal{P}$ is defined that maps the overlap kernel onto the subspace $\mathcal{L}$ of eigenvectors with eigenvalues different from zero (larger than some predefined cut-off value $n_{\sigma}$):
\begin{equation}
    \mathcal{P}_{{\bm {qq'}}}=\sum_{n_l>n_\sigma}\mathcal{U}_{\bm {q}l}\mathcal{U}_{l\bm {q'}}^{\dagger}. 
\end{equation}
Note that both the overlap kernel and the projection operator generally depend on time. The overlap kernel $\mathcal{N}$ and its inverse in the subspace $\mathcal{L}$ read
\begin{equation}
    N_{ik}=(\mathcal{PN})_{ik}=\sum_{n_l>n_\sigma}n_l~\mathcal{U}_{il}\mathcal{U}_{lk}^{\dagger},
\end{equation}
\begin{equation}
    (N^{-1})_{ik}=\sum_{n_l>n_\sigma}(1/n_l)~\mathcal{U}_{il}~\mathcal{U}^{\dagger}_{lk}.
\end{equation}
Similarly, the energy kernel~$\mathcal{H}$ and the mean-field kernel~$\mathcal{H}^{MF}$ are projected onto the subspace $\mathcal{L}$:
\begin{equation}
    H=\mathcal{P}\mathcal{H},~~H^{MF}=\mathcal{P}\mathcal{H}^{MF}.
\end{equation}
With these definitions, the time evolution of the weight functions $f$ in the subspace~$\mathcal{L}$ is determined by the generalized GHW equation \cite{Regnier2019PRC}
\begin{equation} \label{Eq_HW_3}
    i\hbar\dot{f}=N^{-1}(H-H^{MF})f+i\hbar \dot{\mathcal{P}}f.
\end{equation}

\subsection{Collective wave function $g(t)$}
Equation (\ref{Eq_HW_3}) is not a collective Schr\"odinger equation, and the weight function $f_{\bm q}(t)$ is not a probability amplitude of finding the system at the collective coordinate ${\bm q}$. The corresponding collective wave function $g_{\bm q}(t)$ is defined by the transformation \cite{Reinhard1987RPP}
\begin{equation}\label{Eq_f_g}
g=N^{1/2}f,
\end{equation}
where the explicit expression for the square root of the overlap kernel $N$ reads:
\begin{equation}
    (N^{1/2})_{ik}=\sum_{n_l>n_\sigma}\sqrt{n_l}~\mathcal{U}_{il}\mathcal{U}^{\dagger}_{lk}.
\end{equation}
Inserting Eq.~(\ref{Eq_f_g}) into Eq.~(\ref{Eq_HW_3}),
one finally obtains for the time evolution of the collective wave function \cite{Regnier2019PRC}
\begin{equation}
\label{Eq_HW_4}
    i\hbar \dot{g}=N^{-1/2}(H-H^{MF})N^{-1/2}g+i\hbar\dot{N}^{1/2}N^{-1/2}g.
\end{equation}
This equation will be used in the following sections to model small- and large-amplitude collective motion starting from a variety of initial conditions, and with a fully quantum mechanical configuration mixing of TD-DFT trajectories as time-dependent basis states.

\subsection{Observables $\hat{O}$}
The kernel of any observable $\hat{O}$
\begin{equation}
 \mathcal{O}_{\bm {q'q}}=\langle\Phi_{q'}(t)|\hat{O}|\Phi_q(t)\rangle
\end{equation}
can be mapped to the corresponding collective operator $\mathcal{O}^c$:
\begin{equation}
\mathcal{O}^c=N^{-1/2}\mathcal{O}N^{-1/2}.
\end{equation}
The expectation value of the observable $\hat{O}$ in the correlated GHW state is
\begin{equation}
\langle \Psi(t)|\hat{O}|\Psi(t)\rangle =f^\dag \mathcal{O} f=g^\dag \mathcal{O}^c g.
\end{equation}
This expression will be used, for instance, to evaluate the time-dependent multipole moments of the one-body density distribution.

\section{Collective vibrations of $^{\bf{208}}$Pb}\label{sec:208Pb}
\begin{figure}[]
\centering
\includegraphics[width=1.0\textwidth]{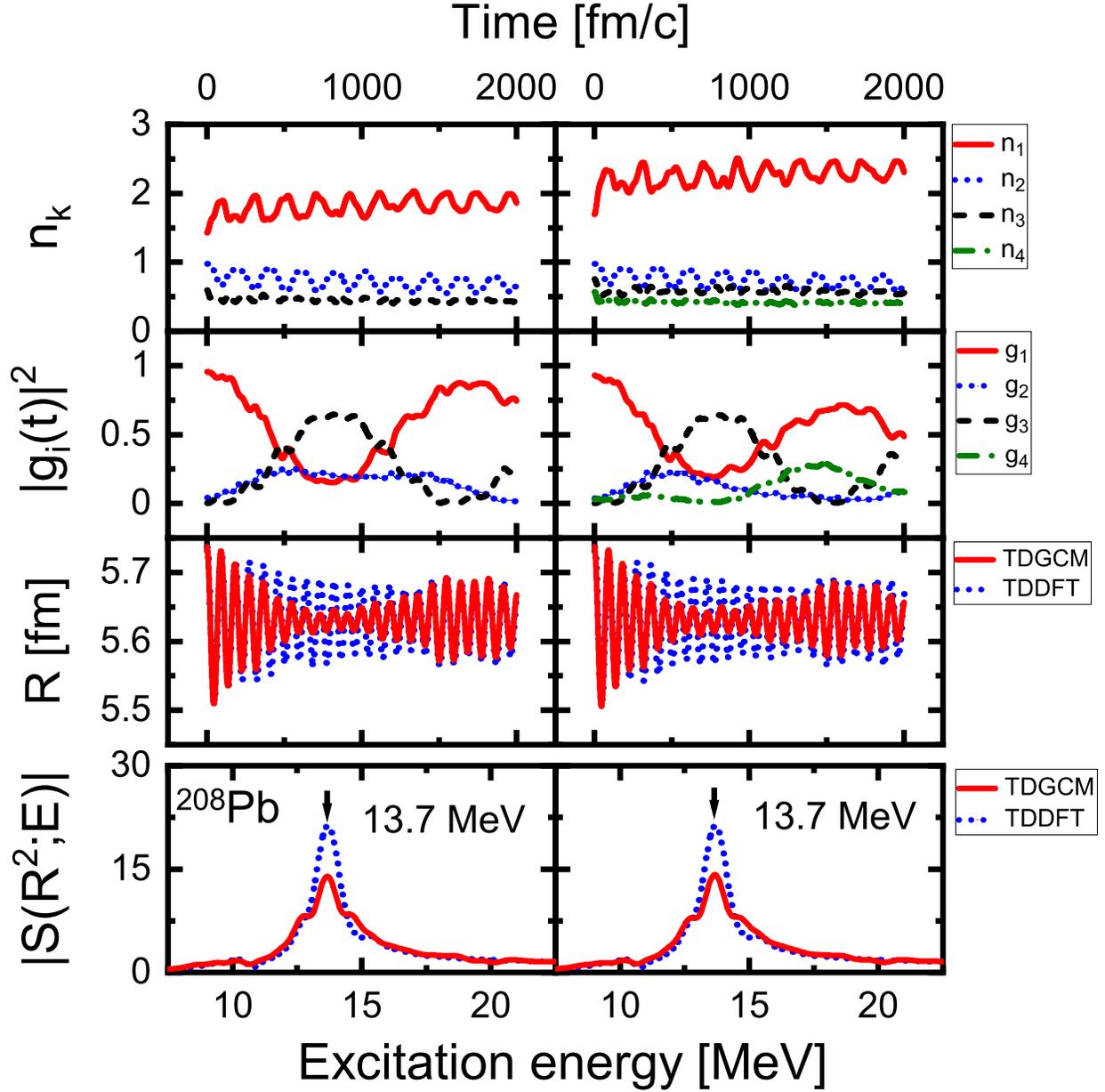}
\caption{Monopole response of $^{208}$Pb, modeled with the TD-GCM. Results obtained with three TD-DFT basis trajectories of initial moments $R_{init} = 5.737$ fm, $\beta_{20, init} = 0.074$, and $\beta_{30, init} = 0.145$ are shown in the left column. A hexadecapole TD-DFT trajectory with $\beta_{40, init} = 0.1$ is included in the time evolution shown in the column on the right. The first row displays the eigenvalues of the overlap kernel, while the square moduli of components of the collective wave function are shown in the second row. In the third row the TD-DFT and TD-GCM radii are shown, and the corresponding strength functions, in units of $10^3~{\rm fm^4/MeV}$, are plotted in the fourth row.}
\label{fig:Pb_mono_2}
\end{figure}

\begin{figure}[]
\centering
\includegraphics[width=0.75\textwidth]{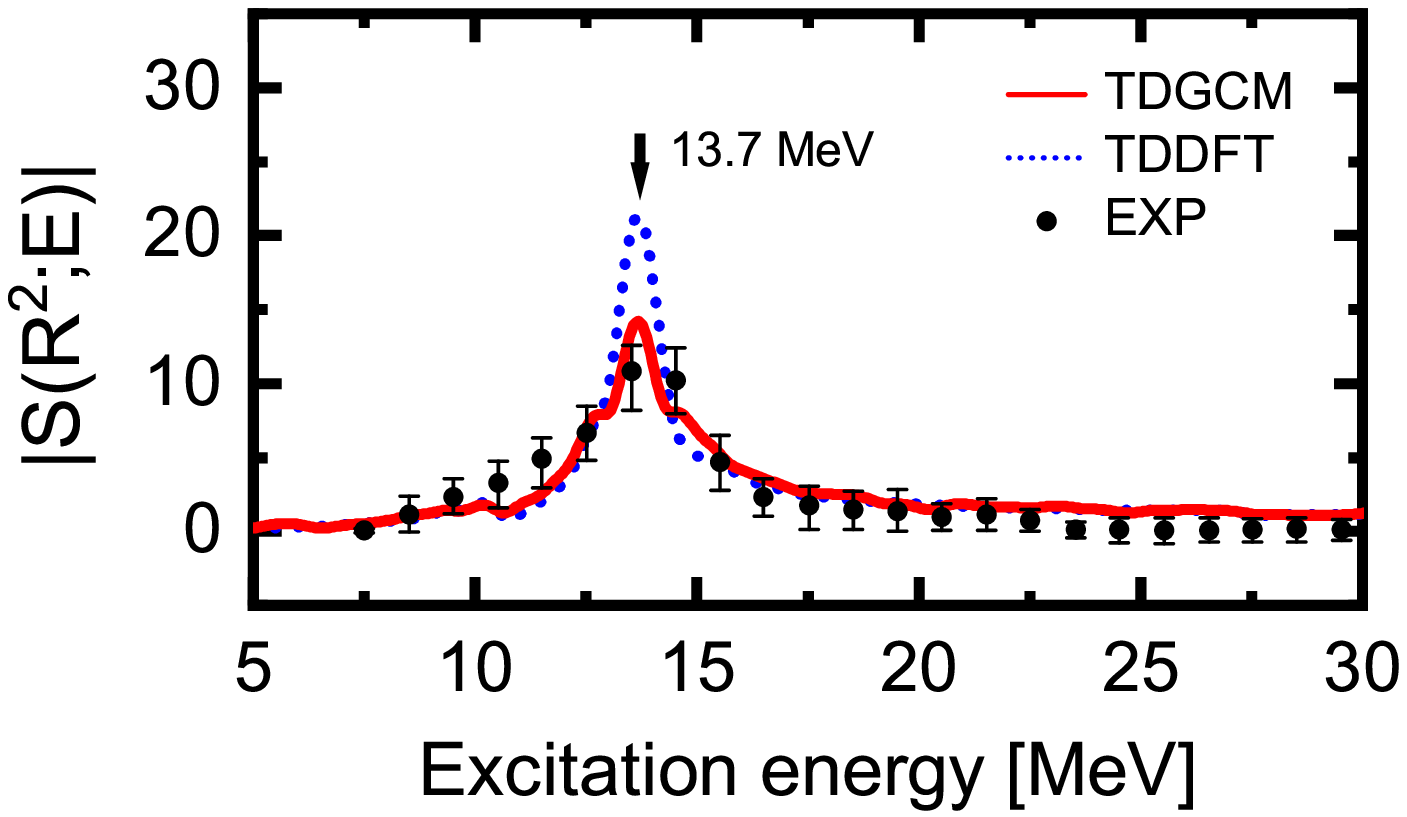}
\caption{The experimental ISGMR strength function in $^{208}$Pb \cite{PatelPhDthesis2014}, in units of $10^3~{\rm fm^4/MeV}$, compared with the results of the TD-DFT and TD-GCM (four basis trajectories) calculations. See text for description.}
 \label{fig:Pb_mono_exp}
\end{figure}

As a first application, we perform an illustrative study of small-amplitude oscillations of the spherical nucleus $^{{208}}$Pb. All calculations are carried out on a lattice in coordinate space~\cite{Ren2017Phys.Rev.C024313,ren19LCS,Li2020} , with the mesh spacing of 1 fm for all directions, and the lattice size is $L_x\times L_y\times L_z=24\times 24\times24~{\rm fm}^3$. The dynamics is determined by the point-coupling relativistic energy density functional PC-PK1~\cite{Zhao2010PRC}. The time-dependent single-particle Dirac equation \eqref{Eq_td_Dirac_eq_BCS}, which provides the TD-DFT Slater determinants as basis states for the GHW equation (\ref{Eq_HW_4}), is solved using the predictor-corrector method, with the time step 0.2~fm/c ($6.67\times10^{-4}$~zs). The initial states for the time evolution are obtained by self-consistent constrained relativistic mean-field (RMF) calculations. The calculated equilibrium binding energy of this spherical nucleus is 1637.97 MeV and the corresponding matter radius is 5.617 fm, in excellent agreement with data.

\begin{figure}[]
\centering
\includegraphics[width=0.75\textwidth]{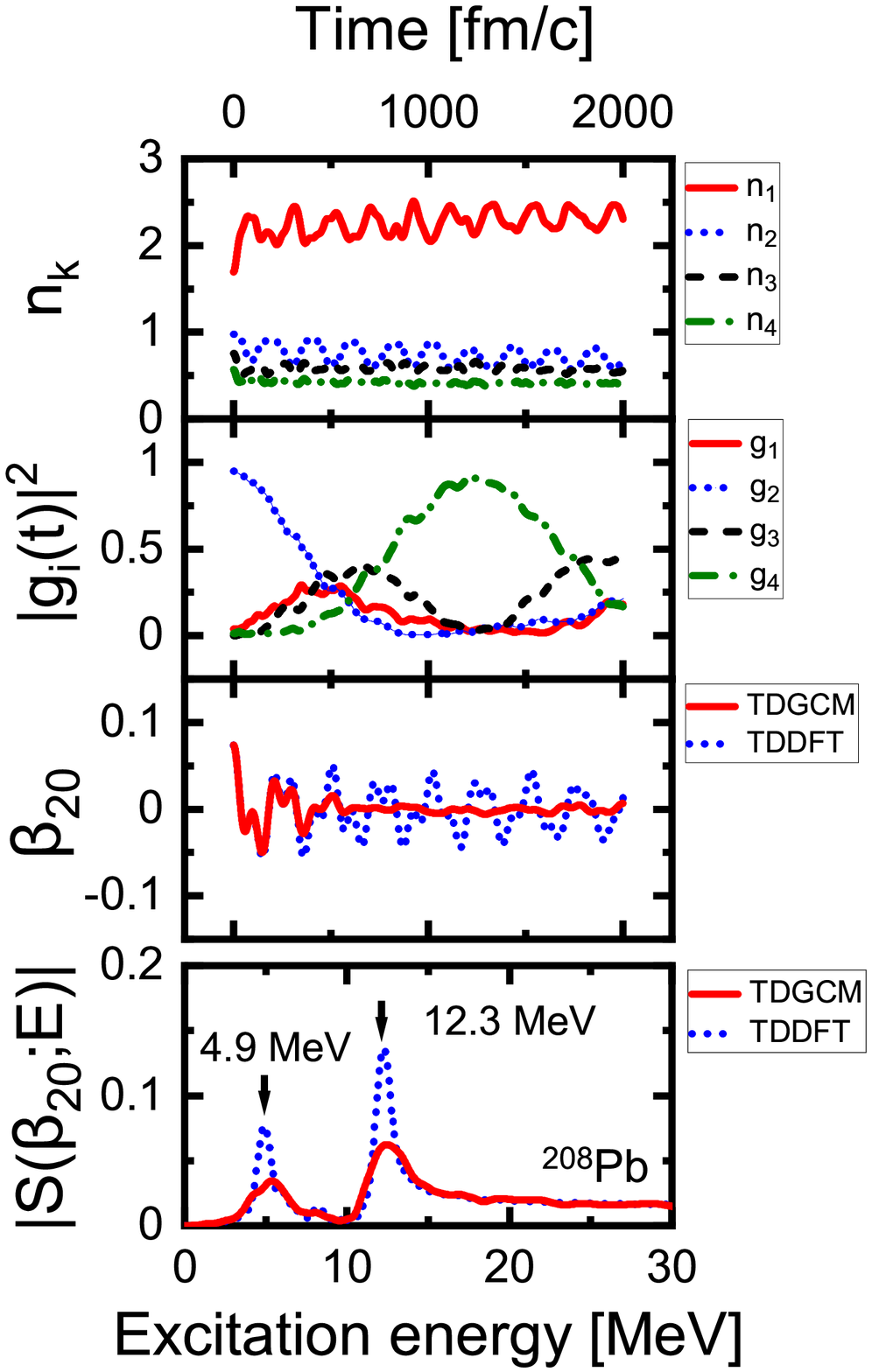}
\caption{Quadrupole response in $^{208}$Pb, modeled using the TD-GCM with four TD-DFT basis trajectories of initial moments $R_{init} = 5.737$ fm, $\beta_{20, init} = 0.074$, $\beta_{30, init} = 0.145$, and $\beta_{40, init} = 0.1$. The top panel displays the eigenvalues of the overlap kernel, and the square moduli of components of the collective wave function are shown in the second panel. In the third panel the time evolution of the mean-field and GCM quadrupole moments is shown, with the corresponding strength functions [${\rm MeV^{-1}}$] plotted in the bottom panel.}
 \label{fig:Pb_quadrupole}
\end{figure}

We consider a basis of TD-DFT trajectories that describe oscillations of different multipolarities and, by using the TD-GCM, form a coherent superposition of these trajectories. Specifically, we combine TD-DFT trajectories that correspond to monopole, quadrupole, octupole, and hexadecapole oscillations. The initial states represent deformation-constrained mean-field solutions. For simplicity, axial symmetry is assumed, and the initially constrained deformations $R_{init} = 5.737$ fm, $\beta_{20, init} = 0.074$, $\beta_{30, init} = 0.145$, and $\beta_{40, init} = 0.1$ all correspond to an energy approximately 6 MeV above the spherical equilibrium minimum.

In Fig.~\ref{fig:Pb_mono_2} we analyze the response to the monopole operator in $^{208}$Pb. The TD-DFT result is already in excellent agreement with the experimental excitation energy of the isoscalar giant monopole resonance (ISGMR). The corresponding strength function of the monopole moment exhibits a single pronounced peak that coincides with the experimental ISGMR at 13.7 MeV. In the left column we compare the TD-DFT result for monopole oscillations with the TD-GCM calculation that combines three TD-DFT basis trajectories (monopole, quadrupole,and octupole) of initial deformations $R_{init} = 5.737$ fm, $\beta_{20, init} = 0.074$, and $\beta_{30, init} = 0.145$. All three eigenvalues of the overlap kernel are large and there is no need for projection of spurious solutions. The collective wave function is initially dominated by the monopole component, but after about 500 fm/c the octupole mode becomes more prominent. The dominant contributions of the monopole and octupole components oscillate with a period of about 1500 fm/c, while the contribution of the quadrupole mode generally remains small for the entire interval of 2000 fm/c. This result is consistent with the fact that the lowest excited level in $^{208}$Pb is the state $3^-$ at 2.61 MeV.

In the TD-GCM calculation illustrated in the column on the right of Fig.~\ref{fig:Pb_mono_2} we have also included, in addition to the monopole, quadrupole and octupole, the hexadecapole TD-DFT trajectory with $\beta_{40, init} = 0.1$. One observes an oscillatory behavior out of phase with the quadrupole component, but both these components are generally much smaller than the monopole and octupole ones. As a consequence, the inclusion of the hexadecapole trajectory produces only a minor effect on the time evolution of the radius, as shown by comparing the two panels in the third row. Generally, the TD-GCM radii exhibit more damping compared to the TD-DFT result, and this is also clearly demonstrated by the corresponding strength functions shown in the fourth row of
Fig.~\ref{fig:Pb_mono_2}. In fact, when compared with the experimental ISGMR strength function \cite{PatelPhDthesis2014} in Fig.~\ref{fig:Pb_mono_exp}, one notices the excellent agreement between the data and the TD-GCM monopole strength function calculated with four basis trajectories. This is not surprising. It is well known that a simple time-dependent mean-field calculation (TD-DFT here) is equivalent to the random phase approximation (RPA) and, therefore, it generally reproduces the excitation energies but not the widths of giant resonances. By allowing for mode coupling, the TD-GCM goes beyond the RPA level and, in principle, should be able to describe the spreading width of resonances. The TD-GCM is, in fact, equivalent to various second-RPA approaches that, in addition to particle-hole ($p-h$) excitations, include also two-particle two-hole states, etc., either directly or through coupling $p-h$ states to selected (multi) phonon states. In particular, it appears that our TD-GCM monopole strength function of $^{208}$Pb is almost identical to the one calculated in a recent study of the nuclear breathing mode of Ref.~\cite{Litvinova2023PRC}, in which, based on a microscopic theory of nuclear response, it has been shown that a parameter-free inclusion of beyond-mean-field correlations of the quasiparticle-vibration coupling type in the leading approximation allows for a simultaneous realistic description of the ISGMR in different mass regions (see Fig. 1 of Ref.~\cite{Litvinova2023PRC}). We have also verified the results by performing a longer TD-GCM calculation, up to 3000 fm/c. While the TD-DFT radius continues to oscillate with only slightly reduced amplitudes, the TD-GCM radius that takes into account the coupling of the monopole, quadrupole, octupole, and hexadecapole modes, is strongly damped after $\approx 2000$ fm/c and the corresponding strength function is very similar to the one shown in Fig.~\ref{fig:Pb_mono_exp}.

\begin{figure}[h!]
\centering
\includegraphics[width=0.75\textwidth]{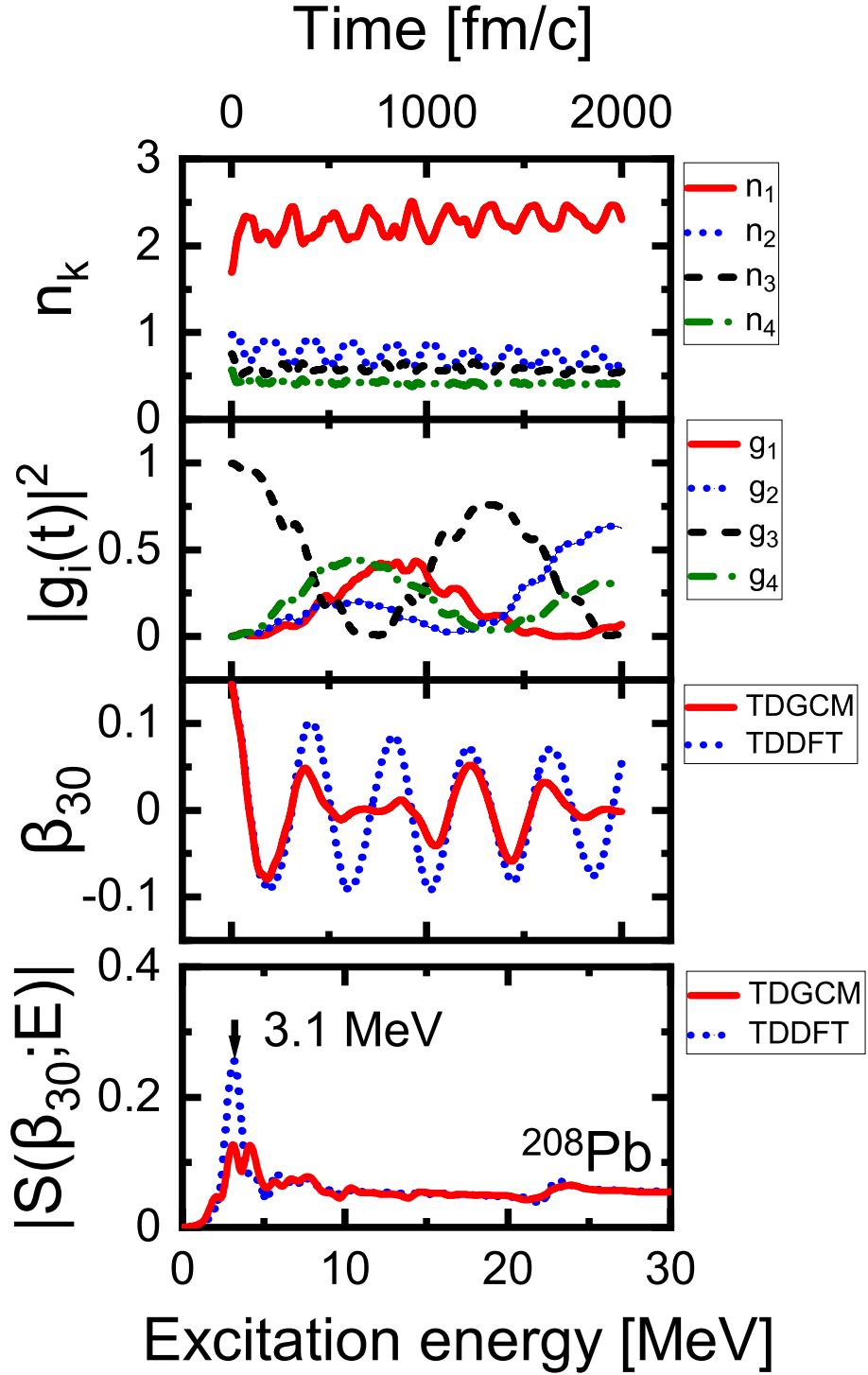}
\caption{Same as in the caption to Fig.~\ref{fig:Pb_quadrupole}, but for the octupole response in $^{208}$Pb.}
 \label{fig:Pb_octupole}
\end{figure}

\begin{figure}[h!]
\centering
\includegraphics[width=0.75\textwidth]{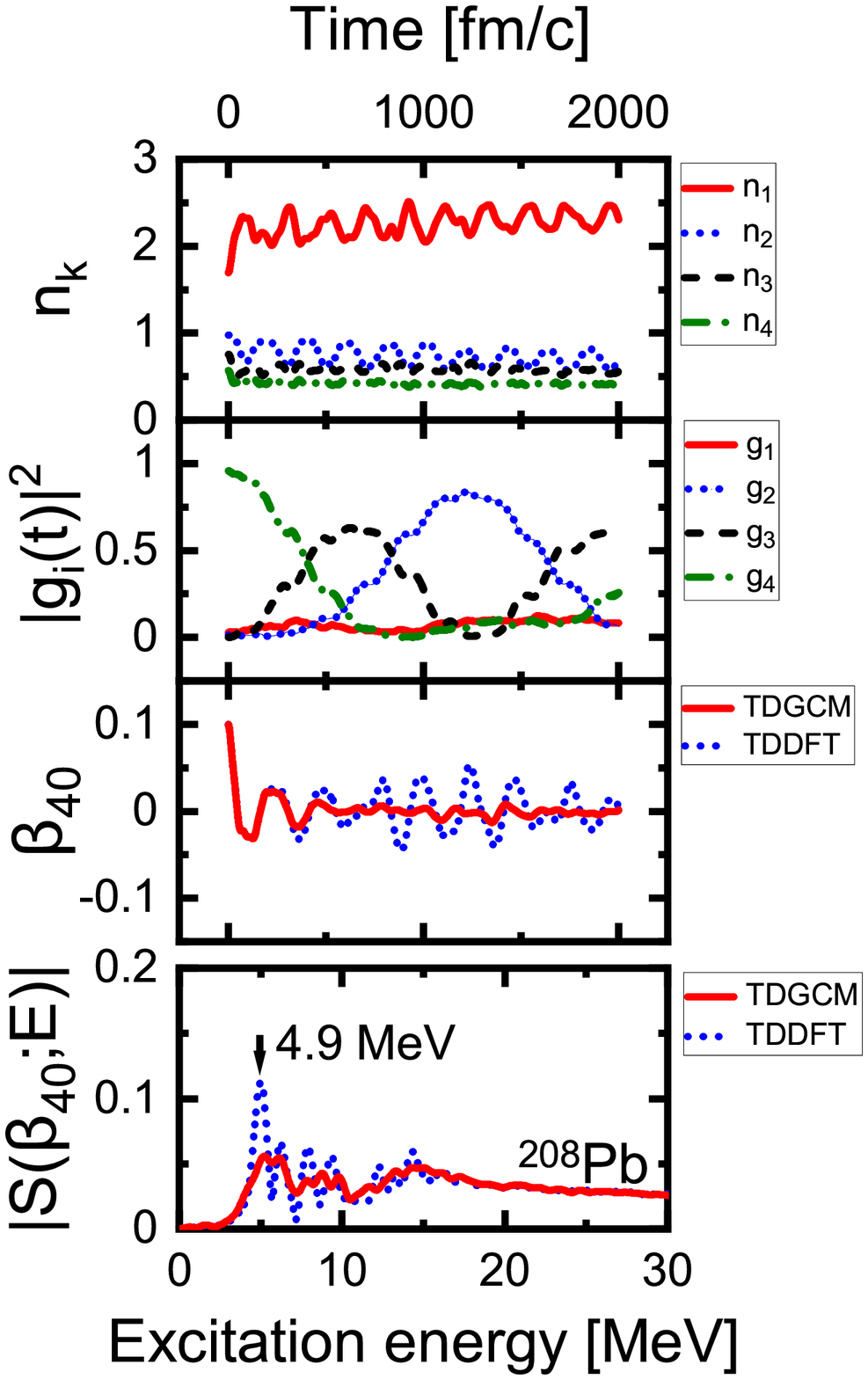}
\caption{Same as in the caption to Fig.~\ref{fig:Pb_quadrupole}, but for the hexadecapole response in $^{208}$Pb.}
 \label{fig:Pb_hexadecapole}
\end{figure}

In Figs.~\ref{fig:Pb_quadrupole}, \ref{fig:Pb_octupole}, and \ref{fig:Pb_hexadecapole}, we display the corresponding TD-DFT and TD-GCM results for the response to the quadrupole, octupole, and hexadecapole operators, respectively. In all three cases the eigenvalues of the overlap kernels are large over the entire interval of time evolution and there is no need to perform projections onto a physical subspace. Generally, the TD-GCM multipole moments exhibit a much more pronounced damping compared to the TD-DFT results, as seen by the time evolution of the quadrupole, octupole, and hexadecapole deformations, and by the corresponding strength functions. In the case of quadrupole oscillations (Fig.~\ref{fig:Pb_quadrupole}), two peaks are clearly identified. The higher lying peak at 12.3 MeV corresponds to the ISGQR, while the one at 4.9 MeV is the low-energy, predominantly $0 \hbar\omega$ quadrupole mode. From the time evolution of the components of the collective wave function, one notices that the strong damping of quadrupole oscillations dominantly arises from the coupling with the hexadecapole and, to a lesser extent, octupole mode. The octupole response is displayed in Fig.~\ref{fig:Pb_octupole}, and here we also note a strong effect of mode coupling in TD-GCM. The TD-DFT peak of the strength function at 3.1 MeV is not far from the position of the lowest experimental $3^-$ state in $^{208}$Pb at 2.61 MeV. Mixing with the other components in the TD-GCM collective wave function, initially in particular with the monopole and hexadecapole modes, leads to a pronounced damping of octupole oscillations. Finally, the hexadecapole case is illustrated in Fig.~\ref{fig:Pb_hexadecapole}. Already the strength function of the TD-DFT hexadecapole deformation parameter exhibits pronounced fragmentation, with the main peak at 4.9 MeV (the lowest experimental $4^+$ state is found at 4.32 MeV). From the time evolution of the components of the collective wave function, we note that this mode strongly mixes with the octupole and quadrupole ones. In fact, after about 500 fm/c the hexadecapole component is almost completely suppressed. The oscillation is strongly damped, as also shown by the corresponding strength function, and this means that hexadecapole oscillations in $^{208}$Pb do not represent a collective mode.

\section{Large amplitude motion}\label{sec_fission}
As emphasized in the introduction, the principal motive to develop the generalized TD-GCM is a description of large-amplitude dynamics, such as the process of induced fission. The idea is to use a basis of, generally non-orthogonal and overcomplete, TD-DFT fission trajectories to build the correlated TD-GCM wave function. TD-DFT automatically includes the one-body dissipation mechanism, but can only simulate a single fission event by propagating the nucleons independently. With the coherent superposition of TD-DFT trajectories in the generalized TD-GCM, fission dynamics is described fully quantum mechanically in an approach that extends beyond the adiabatic approximation of the standard GCM and, at the same time, includes quantum fluctuations. Here we only illustrate the idea with a simple example of two TD-DFT trajectories, while a full analysis will be performed in a forthcoming publication using an implementation of the model that includes pairing correlations, that are essential for a realistic modeling of fission observables.

The example we consider here are fission trajectories of $^{240}$Pu, that were also analyzed in the direct comparison of TD-DFT and TD-GCM of Ref.~\cite{Ren_22PRC}. To be able to follow fission trajectories, the lattice size is $L_x\times L_y\times L_z=20\times20\times60~{\rm fm}^3$, with the mesh spacing of 1 fm for all directions, and the time step 0.2~fm/c. Given the initial single-nucleon wave functions, determined in a mean-field approach with constraints on the collective coordinates in the three-dimensional lattice space, TD-DFT propagates the nucleons independently toward scission. Like in Ref.~\cite{Ren_22PRC}, an axially symmetric two-dimensional collective space of quadrupole $\beta_{20}$ and octupole $\beta_{30}$ deformation parameters of the nuclear density distribution is considered. Since TD-DFT describes the classical evolution of independent nucleons in mean-field potentials, it cannot be applied in the classically forbidden region of the collective space. The starting point of a fission trajectory is usually taken below the outer barrier.

\begin{figure}[]
\centering
\includegraphics[width=0.75\textwidth]{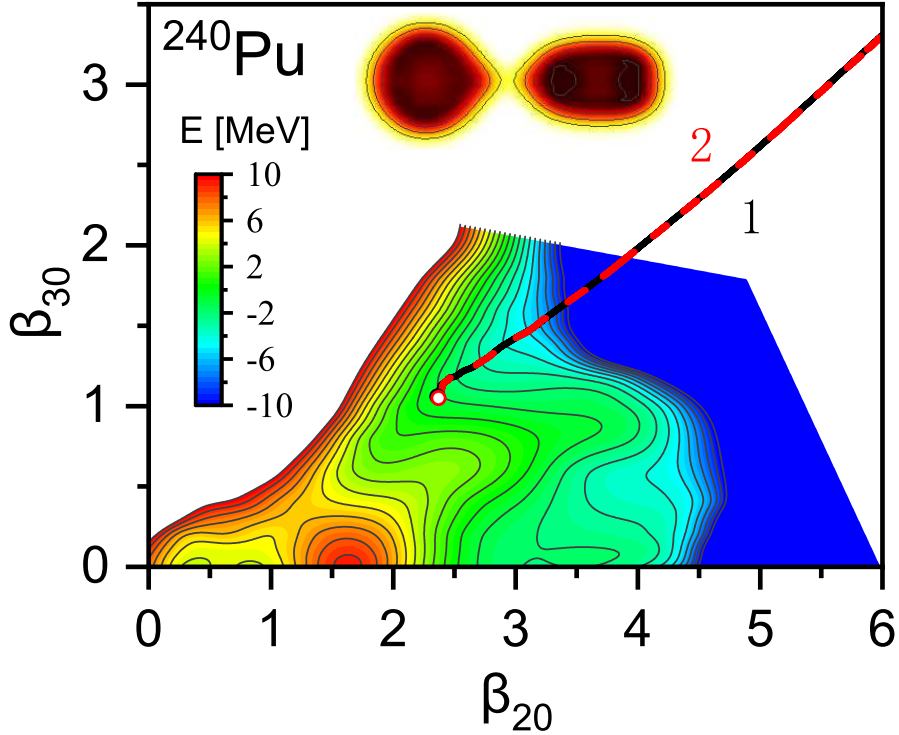}
\caption{TD-DFT fission trajectories from the initial points $\beta_{20}=2.37$ and $\beta_{30}=1.05$, and $\beta_{20}=2.39$ and $\beta_{30}=1.05$, on the deformation energy surface of $^{240}$Pu. The density profile at the instant of scission is shown in the inset.}
 \label{fig:Pu240-paths-close}
\end{figure}

\begin{figure}[]
\centering
\includegraphics[width=0.6\textwidth]{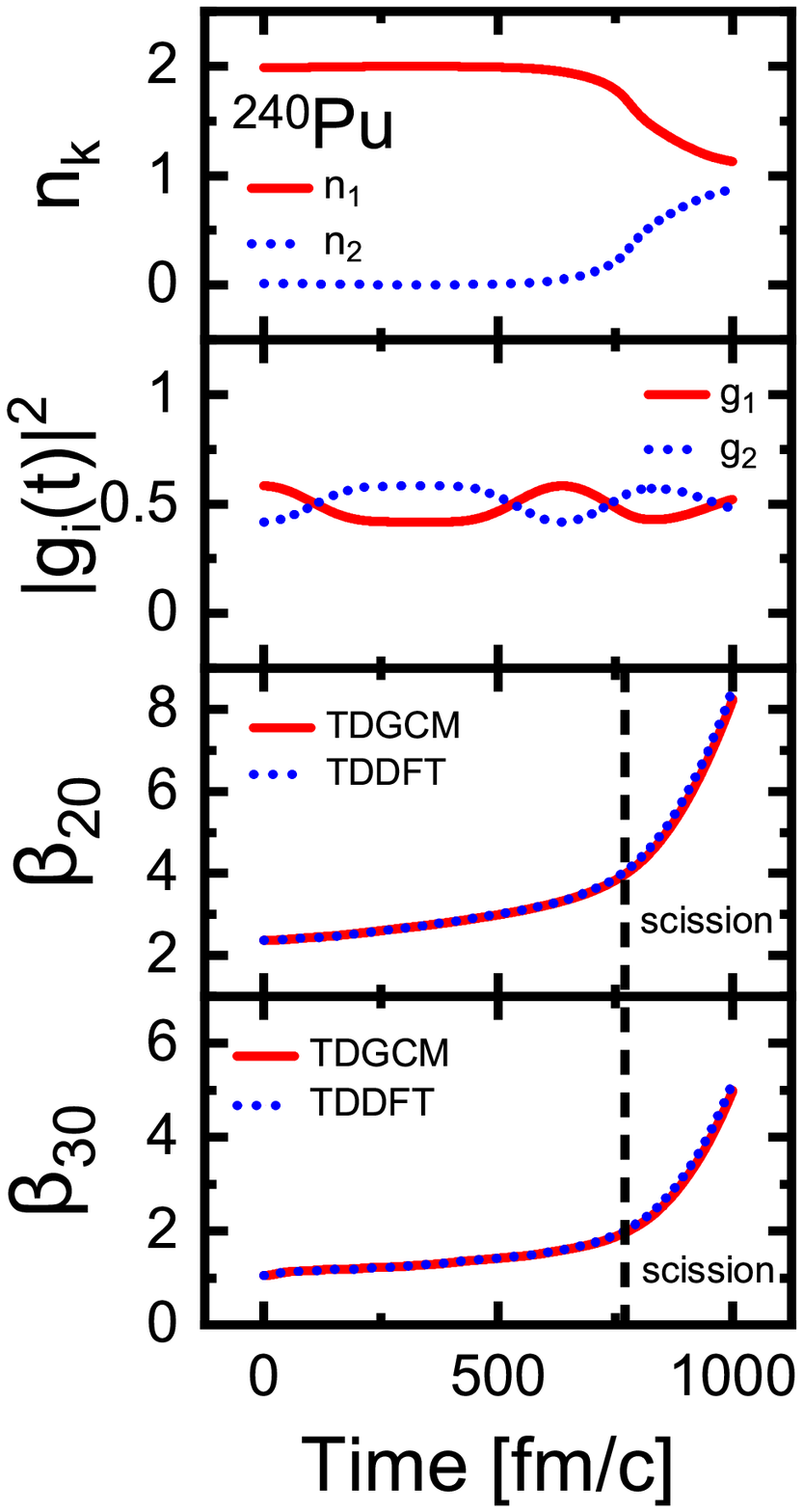}
\caption{Fission of $^{240}$Pu with two TD-DFT trajectories with initial points $\beta_{20}=2.37$ and $\beta_{30}=1.05$, and $\beta_{20}=2.39$ and $\beta_{30}=1.05$, on the deformation energy surface. The top panel displays the eigenvalues of the overlap kernel, and the square moduli of components of the TD-GCM collective wave function are shown in the second panel. The time evolution of the quadrupole and octupole deformations on the way to scission and beyond is compared to the TD-DFT trajectory in the two lower panels. The vertical dashed line denotes the instant of scission.}
 \label{fig:Pu240-tdgcm-close}
\end{figure}

In the first case we examine a superposition of two close-lying fission trajectories on the deformation energy surface of $^{240}$Pu, as shown in Fig.~\ref{fig:Pu240-paths-close}. The initial points are $\beta_{20}=1.05$ and $\beta_{30}=2.37$ for the first trajectory, and $\beta_{20}=1.05$ and $\beta_{30}=2.39$ for the second. Both trajectories lead to scission and remain very close during the time evolution of the fissioning system. In the inset we also show the density profile at the instant of scission. When these trajectories are used as generator states of the generalized TD-GCM, their overlap is large and, therefore, one of the eigenvalues of the overlap kernel is close to 2, while the other vanishes. This is illustrated in the top panel of Fig.~\ref{fig:Pu240-tdgcm-close}, where one also notices that after scission both eigenvalues approach 1 asymptotically with time, which means that the two trajectories become orthogonal. This is because after scission they correspond to distinct pairs of fragments with different particle numbers and, without pairing correlations, automatically become orthogonal. The evolution of the two components of the collective TD-GCM wave function is plotted in the second panel, while the growth of the quadrupole and octupole deformations on the way to scission and beyond is compared to the TD-DFT trajectory in the two lower panels. In this simple example with only two very similar TD-DFT generator states, the evolution of the correlated collective wave function produces a fission event that does not differ from the mean-field result.

\begin{figure}[]
\centering
\includegraphics[width=0.75\textwidth]{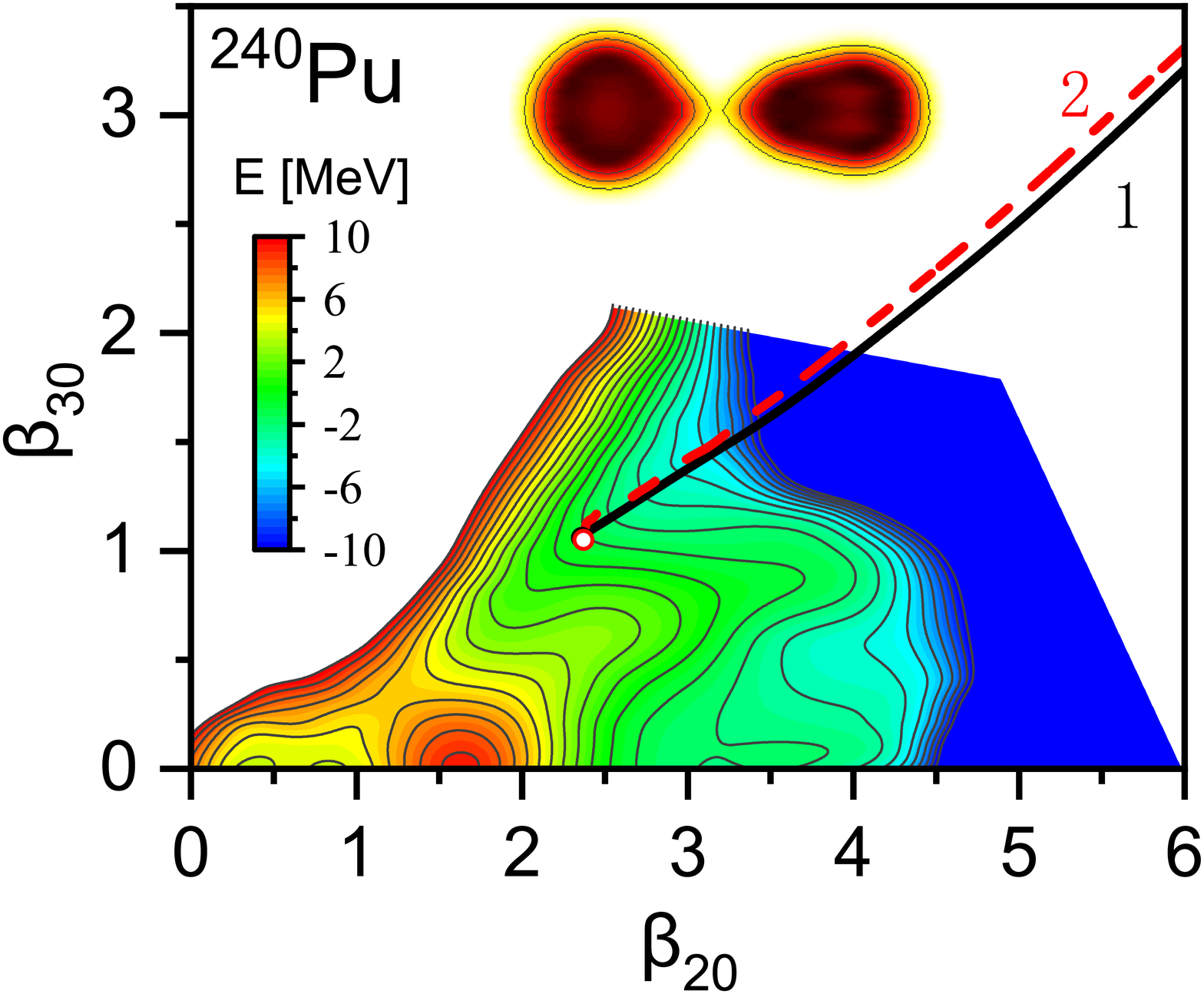}
\caption{Same as in the caption to Fig.~\ref{fig:Pu240-paths-close}, but for the initial points $\beta_{20}=2.36$ and $\beta_{30}=1.06$, and $\beta_{20}=2.37$ and $\beta_{30}=1.05$.}
 \label{fig:Pu240-paths-far}
\end{figure}

\begin{figure}[]
\centering
\includegraphics[width=0.6\textwidth]{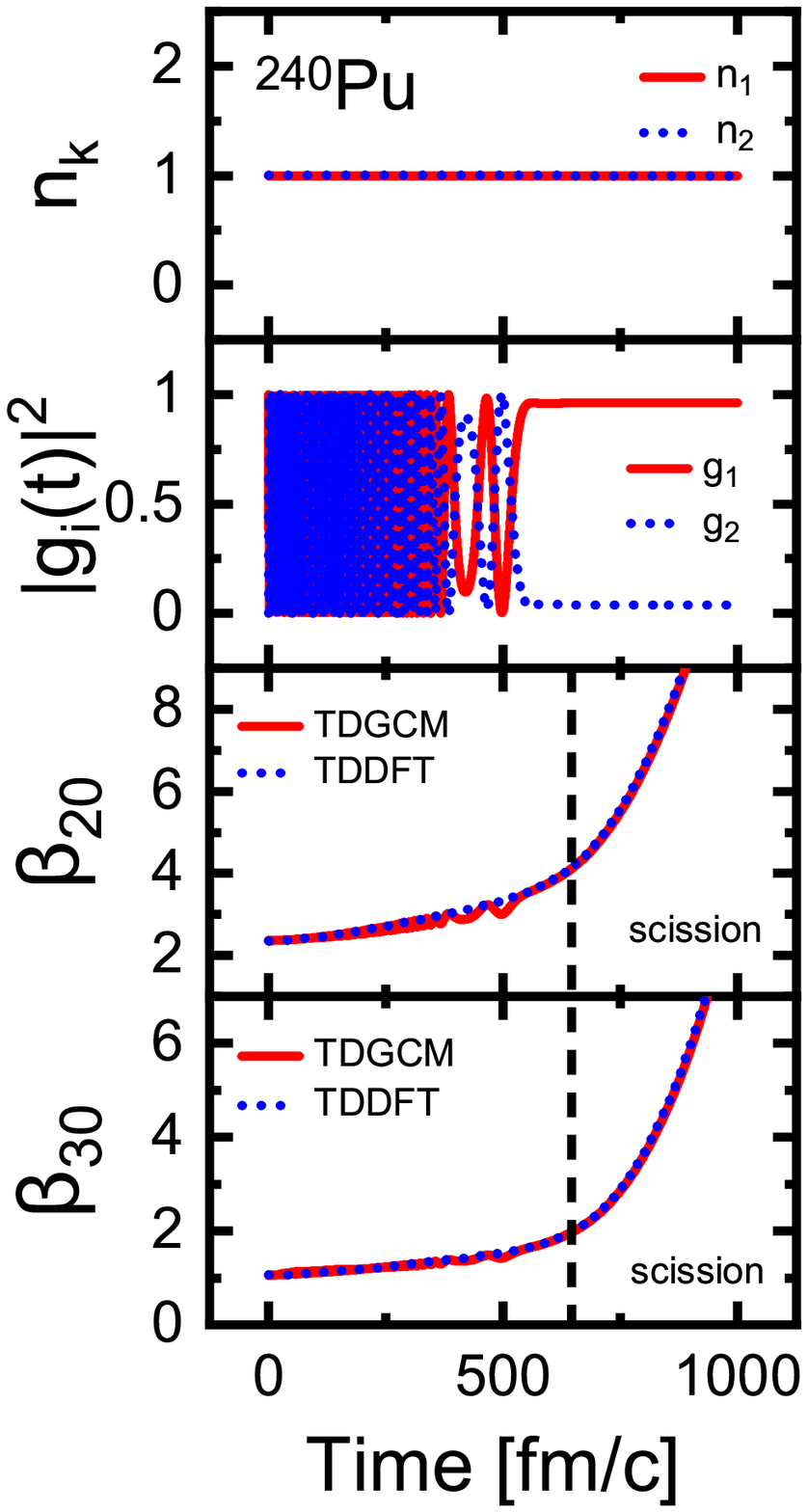}
\caption{Same as in the caption to Fig.~\ref{fig:Pu240-tdgcm-close}, but for the initial points $\beta_{20}=2.36$ and $\beta_{30}=1.06$, and $\beta_{20}=2.37$ and $\beta_{30}=1.05$.}
 \label{fig:Pu240-tdgcm-far}
\end{figure}

The next case illustrates the importance of including pairing correlations and/or finite temperature in the TD-GCM description of fission dynamics. The two TD-DFT trajectories shown in Fig.~\ref{fig:Pu240-paths-far} are not very different from the ones that we have just discussed. They start from almost identical initial points, and also remain very close during the entire time evolution. However, because the trajectories are initially orthogonal and correspond to pure mean-field Slater determinants (the single-particle states are either fully occupied or empty), in the absence of additional correlations they remain orthogonal during the time evolution (top panel of Fig.~\ref{fig:Pu240-tdgcm-far}). Since the trajectories contain essentially the same physical information, the amplitudes of the corresponding components of the collective wave function exhibit very fast unphysical oscillations before scission, and completely separate afterwards (second panel). It appears that the evolution of the deformation parameters is hardly affected (lower two panels) but, of course, no fission observables can be calculated with such a collective wave function. It is thus important to include pairing correlations or finite-temperature occupation factors, to ensure that neighboring trajectories have non-vanishing overlaps.

\section{Summary and outlook}\label{sec_summ}
Nuclear time-dependent density functional theory and the generator coordinated method have been combined in a generalized framework, in which both the generator states and weight functions of the GCM correlated wave function depend on time. This approach goes beyond the usual adiabatic approximation of the time-dependent GCM, and includes the intrinsic one-body dissipation mechanism of TD-DFT. At the same time, it extends the semi-classical TD-DFT to a fully quantum mechanical description of collective dynamics.

For the time-dependent problem, the initial states are obtained as solution of deformation constrained self-consistent mean-field equations. These states are evolved in time by the standard mean-field equations of nuclear DFT. The resulting trajectories form a generally non-orthogonal and overcomplete basis in which the TD-GCM wave function is expanded. The weights, expressed in terms of a collective wave function, obey a time-dependent GCM (integral) equation. In its current implementation, the generalized TD-GCM does not include pairing correlations or finite temperature effects and, therefore, has only limited applicability. In this preliminary paper, the model has been applied to few representative cases of small- and large-amplitude collective motion in nuclei. All calculations have been performed using the relativistic energy density functional PC-PK1.

In the first example we have considered small-amplitude collective oscillations of $^{208}$Pb. The response to the monopole, quadrupole, octupole, and hexadecapole operators has been analyzed. The TD-DFT basis trajectories are initiated using the self-consistent mean-field solutions with constraints on the corresponding deformation parameters, and evolved in time over many periods of oscillations. The Fourier transform of the time-dependent monopole moment determines the corresponding strength function that can directly be compared to data.
Even though already the TD-DFT strength functions yield excitation energies that are in qualitative agreement with data, it is only with the inclusion of mode coupling in the TD-GCM that the spreading widths of resonances can be described. This has been illustrated, in particular, for the ISGMR of $^{208}$Pb, for which both the empirical excitation energy and width are reproduced by the TD-GCM calculation. An interesting feature is also the possibility to follow in time the contribution of the various multipoles in the correlated collective wave functions and, therefore, the method is equivalent to the particle-vibration coupling beyond-mean-field approach.

As an example of large-amplitude motion, we have analyzed the TD-GCM description of induced fission dynamics of $^{240}$Pu. In the simplest case just two fission trajectories can form the TD-DFT basis of the GCM wave function. They can be selected in such a way that their initial overlap is large or small, and the resulting TD-GCM fission dynamics examined. Although schematic, this example indicates the necessity of including pairing correlations and/or finite temperature effects in the basis of TD-DFT fission trajectories.

The implementation developed and applied in the present paper has clearly shown the potential of the generalized TD-GCM framework for a quantitative description of small- and large-amplitude collective motion in nuclei, based on universal EDFs. For a given EDF and pairing interaction, the GCM can be used to simultaneously describe low-energy spectroscopy for various intrinsic deformations, excitation energies and spreading widths of giant resonances, and fission dynamics that includes both dissipation and quantum fluctuations. In the second part of this paper, we will apply an implementation of the generalized TD-GCM that includes pairing correlations and finite temperature effects to a quantitative study of induced fission dynamics. Then, more illustrative examples including different fission systems, such as more neutron-rich Pu isotopes and/or U isotopes~\cite{Rodriguez2014PRC, Rodriguez2014EPJA}, could be studied.

\begin{acknowledgments}
This work has been supported in part by the High-End Foreign Experts Plan of China,
National Key Research and Development Program of China (Contract No.2018YFA0404400),
the National Natural Science Foundation of China (Grants No.12070131001, No.11875075, No.11935003, No.11975031,and No.12141501),
the High-Performance Computing Platform of Peking University,
the QuantiXLie Centre of Excellence [a project cofinanced by the Croatian Government and European Union through the European Regional Development
Fund¡ªthe Competitiveness and Cohesion Operational Programme (Grant No. KK.01.1.1.01.0004)],
and the Croatian Science Foundation under the project ``Uncertainty quantification within the nuclear energy density framework'' (Grant No. IP-2018-01-5987).
\end{acknowledgments}

\bigskip

\section{Appendix}

\subsection{Monopole oscillations of $^{\bf{16}}$O}\label{sec:16O}
In this example, small-amplitude monopole oscillations of $^{16}$O are analyzed. The mesh spacing is 0.8 fm for all directions, and the lattice size is $L_x\times L_y\times L_z=19.2\times19.2\times19.2~{\rm fm}^3$.
The energy density functional is again PC-PK1, and the time-dependent single-particle Dirac equation is solved using the predictor-corrector method, with the time step 0.2~fm/c. As in the case of $^{208}$Pb in Sec.~\ref{sec:208Pb}, the initial states for the time evolution are obtained by constrained mean-field calculations.

\begin{figure}[]
\centering
\includegraphics[width=1.0\textwidth]{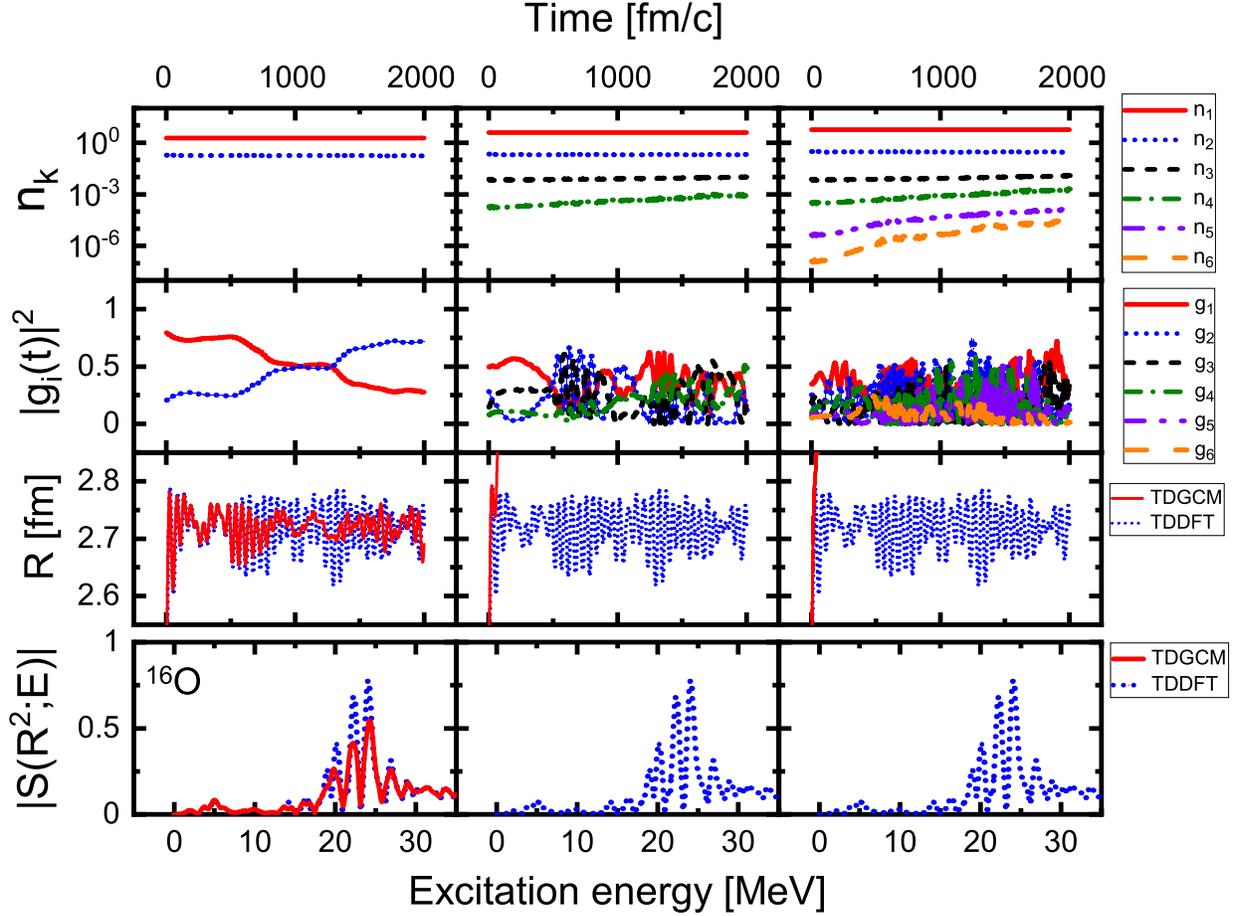}
\caption{Small-amplitude monopole oscillations of $^{16}$O, modeled with the TD-GCM. Results obtained with two TD-DFT basis trajectories of initial radii 2.5 and 2.8 fm; four trajectories of initial radii 2.5, 2.6, 2.7, and 2.8 fm; and six trajectories of initial radii 2.5, 2.55, 2.6, 2.7, 2.75, and 2.8 fm are shown in the left, middle, and right column, respectively. The first row displays the eigenvalues of the overlap kernel, while the square moduli of components of the collective wave function are shown in the second row. In the third row the TD-DFT and TD-GCM radii are shown, and the corresponding strength functions, in units of $10^3~{\rm fm^4/MeV}$, are plotted in the fourth row.}
 \label{fig:O16_w/o_projection}
\end{figure}

The calculated equilibrium binding energy of $^{16}$O is 127.29 MeV and the corresponding matter radius is 2.64 fm. To illustrate the role of projection, we have performed a TD-GCM calculation of monopole oscillations with two, four, and six TD-DFT basis trajectories. The results obtained without projecting the kernels on the space of eigenvectors of the overlap kernel with non-zero eigenvalues are shown in Fig.~\ref{fig:O16_w/o_projection}. The monopole operator is simply $r^2$, and we follow the time evolution of the system up to 2000 fm/c. The three columns compare results obtained with two TD-DFT trajectories of initial radii 2.5 and 2.8 fm; four trajectories of initial radii 2.5, 2.6, 2.7, and 2.8 fm; and six trajectories of initial radii 2.5, 2.55, 2.6, 2.7, 2.75, and 2.8 fm, respectively. In each case, the choice of initial radii corresponds to constrained RMF calculations in which the equilibrium mean-field state is either compressed or expanded. In the first row the eigenvalues of the overlap kernel are displayed on a logarithmic scale, while the square moduli of components of the collective wave function are shown in the second row. The third row displays the evolution of the TD-DFT and TD-GCM radii, and the corresponding strength functions are plotted in the fourth row. In all three examples the TD-DFT radii correspond to a single trajectory with the initial radius of 2.5 fm. Even though one expects that, in a light nucleus such as $^{16}$O, monopole oscillations will exhibit pronounced fragmentation, the TD-DFT strength function  displays a peak structure concentrated in the energy interval between 20 and 25 MeV.

\begin{figure}[]
\centering
\includegraphics[width=1.0\textwidth]{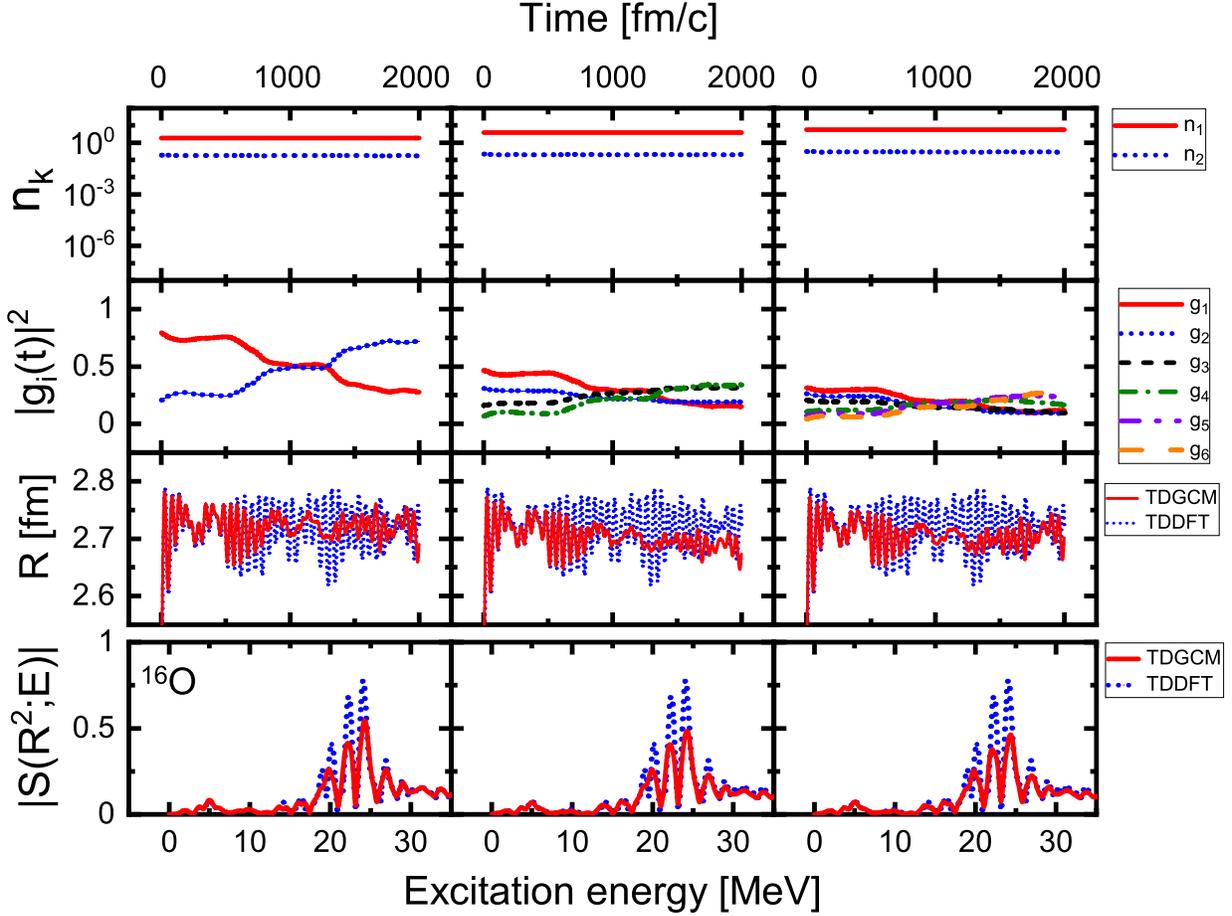}
\caption{Same as in the caption of Fig.~\ref{fig:O16_w/o_projection}, but the results with four and six TD-DFT trajectories are obtained after projecting the kernels onto the subspace of eigenstates of the overlap kernel with eigenvalues larger than $n_\sigma = 0.05$.}
 \label{fig:O16_w_projection}
\end{figure}

By comparing the three columns of Fig.~\ref{fig:O16_w/o_projection}, we note that, without projecting out the spurious eigenvectors of the overlap kernel, stable and realistic TD-GCM results are only obtained in the case with just two TD-DFT trajectories. Both eigenvalues of the overlap kernel are large and constant in time, the evolution of the two components of the collective wave function can be nicely traced in time, and, as a result of this mixing, the signal (TD-GCM radius) exhibits more damping than in the mean-field case. This effect is also clearly reflected in the corresponding strength function, with the main peaks reduced in comparison to the TD-DFT ones. In the second and third column, we note that two and four eigenvalues of the overlap kernel, respectively, are very small. As a result, the components of the collective wave function exhibit very fast unphysical oscillations, and the radius immediately takes unphysical values such that the corresponding strength functions could not be determined.

In Fig.~\ref{fig:O16_w_projection} we display the TD-GCM results obtained with the same choice of TD-DFT trajectories and initial radii but, at each time step, the kernels are projected onto the subspace of eigenstates of the overlap kernel with eigenvalues larger than $n_\sigma = 0.05$. In both cases of four and six trajectories, only two eigenstates are left after projection. Since these are expressed in terms of the original TD-DFT trajectories, the collective wave function has four and six components, respectively. Compared to the calculation without projection, in all cases we are able to follow the TD-GCM evolution of the radius up to 2000 fm/c, and determine the corresponding strength functions. Even though the collective wave function exhibits more mixing in the cases with four and six basis trajectories, in all three examples the signals are very similar, as are the strength functions.
\subsection{Strength function}
Let us assume that a nucleus is initially in its ground state $|\Phi_0\rangle$, with energy~$E_0=0$ at $t=-\infty$, and that an external field $V_{\rm ext}(t)$ is adiabatically switched on:
\begin{equation}
V_{\rm ext}(t)\rightarrow V'_{\rm ext}(t)=\lim_{\epsilon\rightarrow0}V_{\rm ext}(t)e^{\epsilon t}.
\end{equation}
The external field can be expressed in terms of a Fourier transform ($\omega \geq 0$):
\begin{equation}
V_{\rm ext}(t)=\int_{0}^{\infty} [V_{\rm ext}(\omega) F e^{-i\omega t}+V_{\rm ext}^{*}(\omega)F^\dag e^{i\omega t}]~d\omega,
\end{equation}
where $F$ is an arbitrary one-body operator. Accordingly, $V'_{\rm ext}(t)$ reads
\begin{equation}\label{Eq_Vext_lim}
V'_{\rm ext}(t)=\lim_{\epsilon\rightarrow0}\int_{0}^{\infty} [V_{\rm ext}(\omega) F e^{-i(\omega+i\epsilon)t}+V_{\rm ext}^{*}(\omega)F^\dag e^{i(\omega -i\epsilon)t}]~d\omega.
\end{equation}
At time $t$ the nucleus will be in the state
\begin{equation}\label{Eq_Psi}
|\Psi(t)\rangle=|\Phi_0\rangle-i\sum_n e^{-iE_nt}\int^t_{-\infty} dt' e^{iE_nt'}|\Phi_n\rangle\langle \Phi_n|V'_{\rm ext}(t')|\Phi_0\rangle
\end{equation}
in a first-order approximation with respect to $V'_{\rm ext}(t)$. Here, $|\Phi_n\rangle$~and~$E_n$~are the $n$-th excited state and its excitation energy, respectively.
From the expression of Eq.(\ref{Eq_Vext_lim}) for $V'_{\rm ext}(t)$, Eq.(\ref{Eq_Psi}) can be written in the form

\begin{equation}
|\Psi(t)\rangle=|\Phi_0\rangle+\sum_n |\Phi_n\rangle\times\lim_{\epsilon\rightarrow0}\int_{0}^{\infty}[ \frac{V_{\rm ext}(\omega)\langle \Phi_n|F|\Phi_0\rangle}{\omega-E_n+i\epsilon}e^{-i(\omega+i\epsilon) t}
-\frac{V_{\rm ext}^{*}(\omega)\langle \Phi_n|F^{\dag}|\Phi_0\rangle}{\omega+E_n-i\epsilon}e^{i(\omega-i\epsilon) t}]~d\omega.
\end{equation}
The time-dependent expectation value of the operator is defined
\begin{equation}\label{Eq_F_t}
F(t)=\langle\Psi(t)|F^\dag|\Psi(t)\rangle-\langle\Phi_0|F^\dag|\Phi_0\rangle=\lim_{\epsilon\rightarrow0}\int_{0}^{\infty} V_{\rm ext}(\omega)S(F;\omega)e^{-i(\omega+i\epsilon) t}~d\omega+ \ldots,
\end{equation}
where $S(F;\omega)$ is the strength function:
\begin{equation}
S(F;\omega)=\sum_n(\frac{|\langle \Phi_n|F|\Phi_0\rangle|^2}{\omega-E_n+i\epsilon}-\frac{|\langle \Phi_n|F^\dagger|\Phi_0\rangle|^2}{\omega+E_n-i\epsilon}),
\end{equation}
The time evolution of $F(t)$ can also be expressed in terms of a Fourier transform:
\begin{equation}\label{Eq_F_fourier}
F(t)=\int_{0}^{\infty} [ F(\omega)e^{-i\omega t}+F^{*}(\omega) e^{i\omega t}]~d\omega.
\end{equation}
From the equations (\ref{Eq_F_t}) and (\ref{Eq_F_fourier}), one obtains the strength function $S(F;\omega)$:
\begin{equation}
S(F;\omega)=\frac{F(\omega)}{ V_{\rm ext}(\omega)}.
\end{equation}
If initially the nucleus is constrained, as in the present paper, by a mass multipole moment of the density distribution, the external potential $V_{\rm ext}(t)$ takes the form
\begin{equation}
V_{\rm ext}(t) \equiv V_{\rm constr}(t)=\lambda F\theta(-t),
\end{equation}
where $\theta(t)$ denotes the Heaviside step function, $\lambda$ is the constraint parameter, and $F$ is the  operator that corresponds to the specific constraint. Therefore, the Fourier transform $V_{\rm ext}(\omega)$ reads
\begin{equation}
    V_{\rm ext}(\omega)=\lim_{\delta \rightarrow 0} \frac{1}{2\pi}\int_{-\infty}^{\infty} [\lambda \theta(-t)e^{-i(\omega+i\delta)t}]~dt=\frac{\lambda}{2\pi i \omega},
\end{equation}
and, finally, the strength function $S(F;\omega)$ can be evaluated using the expression
\begin{equation}
S(F;\omega)=\frac{2\pi i \omega F(\omega)}{\lambda}.
\end{equation}

\bigskip


%

\end{document}